\documentclass[letterpaper]{aa}
\usepackage{graphicx}
\usepackage{natbib}
\bibpunct{(}{)}{;}{a}{}{,}
\begin{document}

\title{Letter to the Editor:\\
    Convective envelopes in rotating OB stars}

\author{Andr\'e Maeder, Cyril Georgy, Georges Meynet}

\institute{Geneva Observatory CH--1290 Sauverny, Switzerland\\
              email:  andre.maeder@obs.unige.ch\\
              email:  cyril.georgy@obs.unige.ch\\
              email: georges.meynet@obs.unige.ch
               }

\date{Received  / Accepted }

\offprints{Andr\'e Maeder} 
   
\abstract{}{We study the effects of rotation on the outer convective zones of massive stars.}{We examine the effects of rotation on the thermal gradient and on the Solberg--Hoiland term by analytical developments and by numerical models.}{Writing the criterion for convection in rotating envelopes, we show that the effects of rotation on the thermal gradient are much larger and of  opposite sign to   the effect of the Solberg--Hoiland criterion. On the whole, rotation favors convection in stellar  envelopes at the  equator  and to a smaller extent at the poles. In a rotating 20 M$_{\odot}$ star at 94\% of the critical angular velocity, there are two convective envelopes, with the bigger one having a thickness of 13.2\% of the equatorial radius. 
In the non-rotating model, the corresponding convective zone has a thickness of only 4.6\% of the radius.
The occurrence of outer convection in massive stars has many consequences.}{}

\keywords {stars: evolution - convection - rotation }
\titlerunning{Convection in OB Stars}   

\maketitle

\section{Introduction}
It  is generally  considered that the Cowling model applies to massive OB stars: i.e., a convective core
surrounded by a large radiative envelope. However, long since stellar models  have shown that massive stars have an outer  convective envelope  encompassing
several  percent of the stellar radius \citep{Maeder80}. Also, \citet{Langer97} has
shown that an  Eddington factor 
$\Gamma  ={\kappa L}/{(4 \pi c GM)}$ tending toward 1.0
 implies convection.  Our aim  is to show that fast rotation amplifies the size of the convective envelope in OB stars as well as to develop anisotropic convective envelopes.

Various limits can be considered about the effects of rotation and high luminosity on the stellar stability \citep{Langer97,MMVI}: the $\Gamma$--limit, which is the Eddington limit for  $\Gamma \rightarrow 1$; the $\Omega$--limit, which is reached by stars at rotational break--up with a small or negligible 
 effect of the Eddington factor $\Gamma$; the $\Omega \Gamma$-- limit, which applies to stars where both luminosity and rotation play significant roles.
 We   show that not only the stars at the $\Gamma$--limit, but also the stars at  the $\Omega \Gamma$--limit and at the  $\Omega$--limit, have amplified external convective zones.

The occurrence of outer  convective envelopes in OB stars and their  anisotropic structure lead to many astrophysical consequences:
 \begin{itemize}
 \item Convection generates acoustic modes that may allow  asteroseismic observations
 of OB stars.
 \item Convective motions  may play a role in driving mass loss by stellar winds.
 \item For stars close to the critical rotation, convective motions lower 
 the effective break--up velocities.
 \item An outer convective envelope  may make a dynamo and  contribute to some chromospheric activity
   generating an  X--ray emission from OB--stars.
 \item A convective envelope transports   chemical elements and angular momentum.
 \item The occurrence of outer convection may modify the von Zeipel theorem \citep{vZ24}.
 \end{itemize}
A closer investigation is justified. We  start by an analytical approach  (Sect. \ref{analytical}), and finish by  two--dimensional models of 20 M$_{\odot}$  rotating envelopes (Sect. \ref{numerical}).

\section{Convection in rotating stars} \label{analytical}

In a rotating star of mass $M$, luminosity $L$, and angular velocity $\Omega$ (supposed to be shellular, i.e. constant on shells), the total gravity is the sum of the gravitational, centrifugal, and radiative accelerations:
\begin{equation}
 \vec{g_\mathrm{tot}} = \vec{g_\mathrm{eff}} + \vec{g_\mathrm{rad}} = \vec{g_\mathrm{grav}} +
 \vec{g_\mathrm{rot}} + \vec{g_\mathrm{rad}} \;.
 \label{gtot}
\end{equation}%eqn 1.1

\noindent
The vector $\vec{g_\mathrm{eff}}$ has both radial and tangential components, the radial component at colatitude  $\vartheta$ is  
\begin{equation}
g_{\mathrm{eff},r} = - \frac{GM_r}{r^2}\left(1 - \frac{\Omega^2 r^3}{G M_r} \sin^2 \vartheta \right)\; ,
\label{geffr}
\end{equation}

\noindent
where $r$ is the radius at colatitude $\vartheta$.
 The radiative acceleration is directed outward
\begin{equation}
 \vec{g_\mathrm{rad}} = \frac{1}{\rho} \vec{\nabla} P_\mathrm{rad} = \frac{\kappa(\vartheta)\vec{F}}{c} \; ,
 \label{grad}
\end{equation}%eqn 2.6

\noindent 
where $\vec{F}$ is the flux.  On an isobaric surface, $\vec{F}$ is given by the von Zeipel theorem 
\citep{vZ24},
\begin{eqnarray}
\vec{F}  =  - \frac{L(P)}{4 \pi GM_{\star}} 
\vec{g_{\rm{eff}}} \, ,  \; \\
\mathrm{with} \quad M_{\star}(r) = M_r \left( 1 - \frac{\Omega^2}
{2 \pi G \rho_{\rm{m}}}  \right) \; ,
\label{vonZ}
\end{eqnarray}%eqn 2.4

\noindent
and $L(P)$ is the luminosity on the isobar, $\rho_{\rm{m}}$ the internal average density.
In baroclinic stars, there are  other terms  \citep{MMIV}, however they are small  and neglected here. 
The flux is proportional to the effective gravity  $\vec{g_\mathrm{eff}}$. The effective mass  $M_{\star}$ is the mass reduced by  the centrifugal force. Let us note that one  has ${\Omega^2}/(
{2 \pi G \rho_{\rm{m}}}) \approx (4/9) (v/v_{\mathrm{crit}})^2\approx (16/81) \omega^2 $, where 
$\omega=\Omega/\Omega_{\mathrm{crit}}$,
$v$ is the rotation velocity at the level considered and $v_{\mathrm{crit}}=(2/3) [GM/R_{\mathrm{crit}}(\vartheta=0)]^{1/2}$. 
 
\subsection{Effect of rotation on the thermal gradient}  \label{scrot}

Formally the Solberg--Hoiland criterion is to  be considered in a rotating star, as in  Sect. \ref{SH}. However, the radiative  gradient $\nabla_{\mathrm{rad}}$ is also modified by rotation, an effect generally not accounted for in Schwarzschild's criterion.  
 The local   flux 
and  the equation of hydrostatic equilibrium are
\begin{eqnarray}
\vec{F}=-\chi \vec{\nabla} T     \quad \mathrm{and} \quad   
\vec{\nabla}P= \varrho \, \vec{g}_{\mathrm{eff}} \; ,
\label{hydro}
\end{eqnarray}

\noindent
with $\chi= 4 ac T^3/(3 \kappa \varrho)$. Radiation pressure is  included in  
 $P$, the total pressure.
 The local radiative gradient becomes in a rotating star,
 \begin{eqnarray}
 \nabla_{\mathrm{rad}}  = \frac{dT}{dn} \frac{dn}{dP} \frac{P}{T}=
 \,\frac{3}{16 \, \pi \, a \, c \, G } \, \frac{\kappa \, L(P) \, P}
  {M_{\star}(r) \, T^4}   \; , 
 \end{eqnarray}
 
 \noindent
 where the derivatives are computed along a direction $\vec{n}$ perpendicular to the isobars.
 Except $L(P)$, the terms are local and   thus have to be taken at a  given  $(r,\, \vartheta)$. We ignore
 the horizontal thermal gradient and  take  $L(P)$ as constant in the envelope. With (\ref{vonZ})
 and the expression of the Eddington factor $\Gamma$, we get
  \begin{eqnarray}
   \nabla_{\mathrm{rad}}  \, = 
  \,  \frac{\Gamma}{4 \, (1-\beta) \left( 1 - \frac{\Omega^2}
{2 \pi G \rho_{\rm{m}}}  \right) } \; ,
  \label{nablarad3}
  \end{eqnarray}
 
\noindent 
where $\beta= P_{\mathrm{g}}/P$  is the ratio of gas pressure to the total pressure, thus 
$P/(aT^4) = 1/[3(1-\beta)]$.
  The adiabatic gradient $\nabla_{\mathrm{ad}}$ is
 \begin{eqnarray}
 \nabla_{\mathrm{ad}}= 
 \frac{8-6 \beta}{32-24 \beta -3 \beta^2} \; .
 \label{ad}
 \end{eqnarray}
 
 \noindent
 As $T$ varies with $\vartheta$, $\beta$ also varies with colatitude, and we write $\beta(\vartheta)$.
That $\beta(\vartheta)$ is higher at the equator favors equatorial convection. 
The  criterion for convective instability $\nabla_{\mathrm{rad}} > \nabla_{\mathrm{ad}}$
 becomes
 \begin{eqnarray}
 \frac{\Gamma(\vartheta)}{\left( 1 - \frac{\Omega^2}
{2 \pi G \rho_{\rm{m}}}  \right)} \, > \, 4 \, [1-\beta(\vartheta)] \, \nabla_{\mathrm{ad}}  \; ,
 \label{Srot}
 \end{eqnarray}

 \noindent
 where the $\vartheta$--dependence of $\Gamma$ comes only through $\kappa(\vartheta)$ \citep{MMVI}.
 Eq. (\ref{Srot}) has various interesting consequences:
 \begin{itemize}
 \item  In the absence of rotation, expression (\ref{nablarad3}) is equivalent to Langer's result \citep{Langer97}. The  right--hand side of Eq. (\ref{Srot})  is always  smaller than
$1.0$, thus if $\Gamma \rightarrow 1$, the criterion is satisfied. Convection is present in layers close to  the Eddington limit.
\item  In a rotating star,  inequality (\ref{Srot}) is more easily satisfied.
 Thus,  rotation favors convection in stellar  envelopes. 
 \item The occurrence of convection depends on both $\kappa(\vartheta)$ and $\beta(\vartheta)$. Equatorial ejection is always favored, even for electron scattering opacity caused by the higher $\beta$.
 \item When the centrifugal force can be derived from a potential
(conservative case), the temperature and density are constant
on isobars and so that $\Gamma$ and $\beta$ are also constant on isobars. In that case,
rotation also favors convection as can be seen from Eq. (9).
\end{itemize}

\begin{figure}
\centering
\includegraphics[angle=0,width=9.0cm]{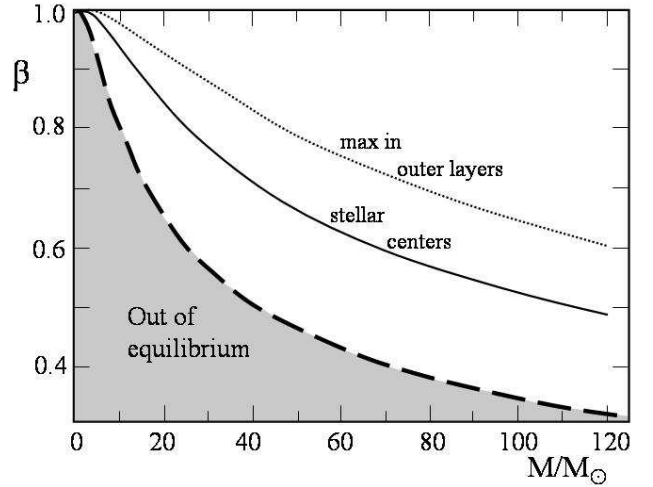}
\caption{The ratio $\beta= P_{\mathrm{gas}}/P $ as a function of the
stellar masses on the zero--age main--sequence with $Z=0.02$. The continuous line shows the value at the stellar centers, the
dotted  line  the maximum value of $\beta$ inside these models. The long--dashed line indicates the minimum value of $\beta$ permitted by equilibrium conditions \citep{Chandra84,Mitalas97}.}
\label{Bbeta}
\end{figure} 
  
  One can wonder whether rotating stars that are neither at the critical nor  at the Eddington limit may develop a convective envelope. The ratio $\beta$ decreases with mass (Fig. \ref{Bbeta}), while the Eddington factor $\Gamma$ increases,
  e.g. $\Gamma= 2.5 \times 10^{-5}, 0.0047, 0.021, 0.098, 0.239, 0.343, 0.544$
  for 1, 5, 9, 20, 40, 60, and 120 $M_{\odot}$ stars on the ZAMS. The
 parameter  $\beta$ is at a minimum in the stellar centers, reaches a maximum in the envelope, and is zero at the stellar surface.  One has the following relation between  the maximum $\beta$--values  in the outer layers and $\Gamma$,
\begin{eqnarray}
\beta \, = \,  1.0 - s \, \Gamma \quad \quad \mathrm{or} \quad \Gamma\, = \, \frac{1}{s} (1-\beta) \, , \quad
\label{betag}
\end{eqnarray}

\noindent
with $s=0.72 \pm 0.01$  between 20 M$_{\odot} $ and 120 M$_{\odot}$.
Using Eq. (\ref{betag}), one eliminates $\Gamma$ from  criterion (\ref{Srot}) and gets
 \begin{eqnarray}
  \frac{1}{s\left( 1 - \frac{\Omega^2}
{2 \pi G \rho_{\rm{m}}}  \right)} \; > \; 4 \; 
 \frac{8-6 \beta}{32-24 \beta -3 \beta^2} \; .
 \end{eqnarray}
 
\noindent
This  relation indicates above which value of $\omega$ there is convection for a given value of 
$\beta$ at the maximum (for lower $\beta$ in the envelope, the inequality  is evidently satisfied more easily). For example, for $\beta \rightarrow 1.0$, the inequality becomes
$\frac{\Omega^2}
{2 \pi G \rho_{\rm{m}}} > 0.132$ or $(v/v_{\mathrm{crit}})> 0.54$.
 This is an approximation 
owing to the simplified relation adopted and  because the various  parameters also vary with depth.
However, it shows  that rotating  massive stars not even at the critical limit may have 
 enhanced convection.

\subsection{The Solberg--Hoiland  criterion}  \label{SH} 

A fluid element displaced in a rotating star is also subject to the restoring effect of angular momentum conservation. This leads to the Solberg--Hoiland criterion for stability
\citep{KW90}, which is (for constant 
mean molecular weight $\mu$)
\begin{eqnarray}
\nabla_{\mathrm{ad}}-\nabla_{\mathrm{rad}}+\nabla_{\Omega} \sin \vartheta > 0  \\ [2mm]
\mathrm{with} \quad \nabla_{\Omega}= \frac{H_P}{g_{\mathrm{grav}} \delta} \, \frac{1}{\varpi^3} \frac{d(\Omega^2 \varpi^4)}{d\varpi} \; ,
\end{eqnarray}

\noindent
where $\varpi= r \sin \vartheta$ is the distance to the rotation axis and $\delta = -
  \left({\partial \ln \varrho}/{\partial \ln T}\right)_{P }$.  The quantity $\nabla_{\Omega}$
  depends on the distribution of the specific angular momentum $j = \varpi^2 \Omega$, which results from  transport processes. As $j$ decreases outward, $\nabla_{\Omega}$ generally has a stabilizing effect. Let us consider the  two extreme cases for $\Omega(r)$:
 
 \noindent 
  -- 1. {\emph{Constant specific angular momentum:}}  a distribution $\Omega \sim r^{-2}$ may result from the Rayleigh--Taylor instability. This distribution is also sometimes  considered in convective regions,
  with the argument that  the plumes  rapidly redistribute the angular momentum. If so, $\nabla_{\Omega}=0 $, and one is  brought back to Schwarzschild's criterion.
  
  \noindent
  --2. {\emph{Constant angular velocity:}} this  assumption is also used in convective regions, with the argument that  turbulent viscosity  favors solid  rotation. If so, $\nabla_{\Omega}$ simplifies 
  to
  \begin{eqnarray}
  \nabla_{\Omega} = 4 \frac{\Omega^2 }{g_{\mathrm{grav}}} \, \frac{H_P}{\delta}= \frac{4 \, \Omega^2}{g_{\mathrm{grav}} \,\varrho }  \, \frac{P}{ g_{\mathrm{eff}} \delta} \; .
  \end{eqnarray}
  
  \noindent
  We can simplify this expression further. 
  In the outer layers, as long as  $\kappa \approx $ const.  and $ g_{\mathrm{eff}} \approx$ const, at an optical depth $\tau$  one has  $P \, \approx \, ({g_{\mathrm{eff}}}/ {\kappa})\, \tau$. This gives
  \begin{eqnarray}
  \nabla_{\Omega} \, \approx \, 4 \, \frac{\Omega^2}{g_{\mathrm{grav}} \, \varrho \, \kappa \delta} \, \tau \, \approx
  4 \,  \left(\frac{\Omega^2 R^3}{G \, M}\right) \, \left[\frac{\tau}{\varrho \, \kappa \, R \delta} \right] \; .
  \end{eqnarray}
  
  \noindent
  The term in the first parenthesis is $\omega^2$, while that in square brackets is just the ratio $(R-r)/R$
  (assuming $\delta=1$), which  is small in the envelope. The Solberg--Hoiland criterion becomes 
  in this approximation,
   \begin{eqnarray}
 \frac{\Gamma}{\left( 1 - \frac{\Omega^2}
{2 \pi G \rho_{\rm{m}}}  \right)} \, > \, 4 \, (1-\beta) 
\left(
\nabla_{\mathrm{ad}} 
 + \omega^2 \left[\frac{R-r}{R}\right] \sin \vartheta\right),
 \label{SHrot}
 \end{eqnarray}

\noindent
where as above the various quantities  are local ones.
At low rotation, the Solberg--Hoiland term $\nabla_{\Omega}$ is negligible with respect to the other terms.
At high rotation for constant $\Omega$, it is not negligible, but  in general smaller than the other terms
because the convective zone
lies very close to the surface and the term
$(R-r/R)$ is small. 

Since the actual rotation laws are likely between the two extreme cases $\Omega(r)=$ const. and $\Omega \sim r^{-2}$,  
 we conclude that the main effect of rotation on convection   in stellar envelopes is not the inhibiting effect due to the Solberg--Hoiland criterion, but  the effect of rotation on the thermal  gradient (Eq. \ref{Srot}), which enhances convection. 

\section{Numerical models}   \label{numerical}

We do some 2-D models 
 of the outer regions of a 20 M$_{\odot}$  fast--rotating star with $X=0.70$ and $Z=0.020$ (Fig. \ref{P94}).
 At each latitude we integrate the equations of the structure for the corresponding
 effective gravity and $T_{\mathrm{eff}}$ of the Roche model of the given rotation, also taking the effect of the reduced mass into account.
In the envelope, we suppose that $\Omega$ is a constant
as a function of depth (the problem is  conservative).
 We first consider only the effect of rotation on the thermal gradient  and then the complete Solberg--Hoiland criterion to see the differences.

\subsection{Effects of rotation on the thermal gradient}  \label{therm}

Without rotation, a 20 $\mathrm{M}_\odot$ model at the end of the MS evolution has two outer convective zones. The first one is very close to the surface and is due to an increase of the opacity caused by partial He ionization. It extends from $r/R = 0.992$ to $0.999$, i.e.  only 0.7\% of the radius, and contains a very small fraction of the total stellar mass ($2.5\cdot 10^{-9}$). The second one is deeper, between $r/R = 0.915$ to $0.962$ (4.7\% of the total radius), and contains  a fraction  $7.4\cdot 10^{-7}$ of the total mass. Both convective zones are associated with  opacity enhancements.

\begin{figure}
\includegraphics[angle=00,width=8.8cm]{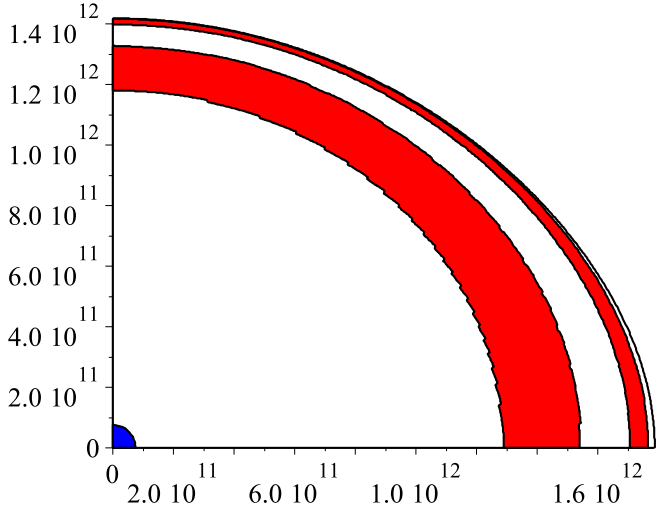}
\caption{2--D model of the  external convective zones  and of the convective core (dark areas) in a model of 20 M$_{\odot}$ with $X=0.70$ and $Z=0.020$ at the end of MS evolution with fast rotation ( $\omega=\Omega/\Omega_{\mathrm{crit}}=0.94$). The axes are in units of cm.}
\label{P94}
\end{figure}

For fast rotation, with a ratio $\Omega/\Omega_{\mathrm{crit}}=0.94$, where $\Omega_{\mathrm{crit}}$ is the 
critical  angular velocity, the convective  layers are  shown in  Fig. \ref{P94}.  They are more extended  
than without rotation. The  thin upper convective zone
extends from $r/R = 0.987$ to $0.999$ at the pole, i.e., over 1.2\% of the stellar radius, and it contains  $1.3\cdot 10^{-8}$ of the stellar mass. The deeper convective zone  covers the region between $r/R = 0.836$ to $0.936$ (i.e. 10.0\% of the polar radius), and its  mass fraction is $2.8\cdot 10^{-6}$. At the equator, we have the following sizes for the two  convective layers: between $r/R = 0.958$ and $r/R = 0.988$  for the first one (3.0\% of the equatorial radius) and between $r/R = 0.727$ and $r/R = 0.862$ for the deeper one (13.5\% of the radius). The included masses are the same as at the pole.

In agreement with  Sect. \ref{scrot}, convection is more extended in the rotating model than in the non--rotating one. Fig. \ref{P94} shows that contrarily to the classical Cowling model, massive rotating 
stars have a large--size convective envelope.
Interesting is that, if we look at the structure of the envelope of the star as a function of the pressure, we see that this structure is independent of the colatitude. 
This is expected since we have supposed $\Omega$ constant in the
envelope. In that case, as recalled above, $\Gamma$ and $\beta$ are constant
on isobars and the extensions of the convective zones, expressed
in term of {\it differences of pressure} between the bottom and the top
of the convective zone are the same at the pole and at the equator.
The {\it spatial} extensions are, however, greater in the equatorial region
than in polar ones.
This comes from the variations in the
spatial gradients of the pressure and temperature with the colatitude imposed by the
hydrostatic equilibrium (gradients of pressure have to balance $\rho g_{\rm eff}$).
Another result of the constancy of pressure and temperature (and thus density) on an isobar is that the radiative gradient is also constant on this isobar, except for the change in the effective mass
 as given 
 by  $M_*$, which is lower than $M$ in rotating stars. Thus, for  rotating stars  the radiative gradient is larger  and convection is favored not only at the equator, but at each colatitude, compared with the non--rotating model.

The mass loss rate in the considered model is $6.2 \times 10^{-7}$ M$_{\odot}$ yr$^{-1}$. This means that within one year about 4 times all the matter in the thin convective zone is lost in the stellar winds! Thus, the matter carried by  the winds is continuously passing through the superficial convective zones in a dynamical process.

\subsection{Solberg--Hoiland  criterion }

\begin{figure}
\includegraphics[angle=00,width=9 cm]{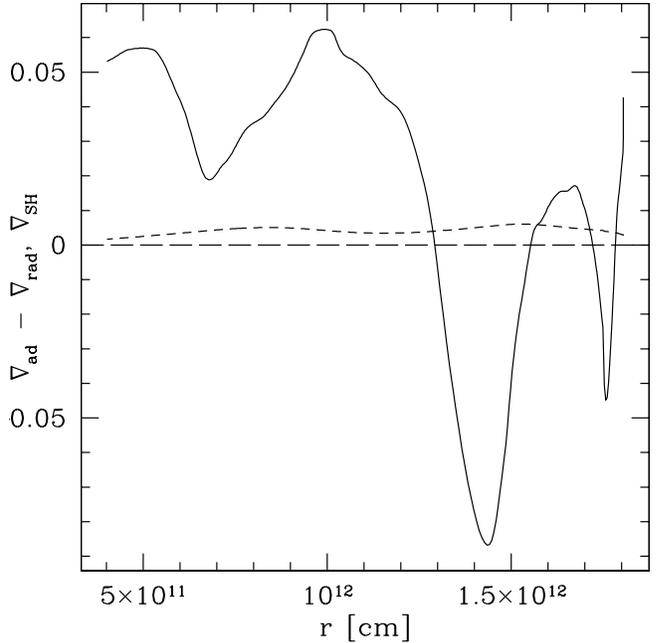}
\caption{$\nabla_\mathrm{ad}-\nabla_\mathrm{rad}$ (solid line) and $\nabla_\Omega \sin\vartheta$ (short--dashed line) at the equator in a  $20\  \mathrm{M}_\odot$ model with $\Omega/\Omega_{\mathrm{crit}}=0.94$ at the end of the MS.
The long--dashed line indicates the zero level.}
\label{Nab94}
\end{figure}

We also computed the envelope structure with the Solberg--Hoiland criterion for  convection. The values of $(\nabla_\mathrm{ad} - \nabla_\mathrm{rad})$ and $\nabla_\Omega \sin\vartheta$ are shown in Fig. \ref{Nab94} for a $20\  \mathrm{M}_\odot$ model at the equator, {\it i.e.} where the effect of the Solberg--Hoiland criterion is the strongest. The solid line shows the difference between the adiabatic and the radiative gradient
(identical to the values obtained in the model computed with the Schwarzschild criterion).
The short-dashed line shows the Solberg-Hoiland term. The limits
of the convective zones when the Soilberg-Hoiland criterion is used
are inside the regions where  $(\nabla_\mathrm{ad} - \nabla_\mathrm{rad})$ is negative. Indeed,
the limits are where $(\nabla_\mathrm{ad} - \nabla_\mathrm{rad})+\nabla_\Omega \sin\vartheta=0.$
Convective zones are thus smaller than those obtained with the Schwarzschild criterion.
We find the following values
for the extension of the convective zones at the equator when the Solberg-Hoiland criterion is used: between $r/R = 0.960$ and $r/R = 0.988$ for the superficial one (2.8\% of the equatorial radius) and between $r/R = 0.727$ and $r/R = 0.859$ for the deeper one (13.2\% of the total radius), i.e. values slightly smaller than but very similar to those obtained in the model computed
with the Schwarzschild criterion.
This numerical example confirms that
the Solberg--Hoiland  term $\nabla_\Omega \sin\vartheta$ has a very limited influence in stellar envelopes, as discussed in Sect.~2.3.

\section{Conclusions}

In stellar envelope of rotating stars, the effects of rotation on the thermal gradient arestronger and with the opposite sign with respect to the Solberg--Hoiland criterion, so that rotation favors convection
instead of inhibiting it. The increase of the convective zone occurs mainly 
at the equator  and  also a bit at the poles. In a fast--rotating 20 M$_{\odot}$ 
Pop I star, there are  two  equatorial zones covering a total of 16\% of the stellar radius at the equator.
 
There are several consequences of thees results to be examined in future.
The outer convective motions may   lower the escape velocity as well as the critical rotation velocity.
The matter accelerated in the winds continuously goes through the convective zone in a dynamical process, suggesting  that convection plays a role in accelerating the stellar winds
and in producting the clumps in the winds.
The convective pistons generate acoustic waves of  periods of several hours to a few days.
The density is very low, and it is thus likely that convection injects oscillations into the wind
rather than into the interior.

\bibliographystyle{aa}
\bibliography{9007}

\end{document}